\documentclass[12pt]{article}

\usepackage{latexsym,amsmath,amssymb,theorem,epsfig}

\usepackage{cite}

\newcommand \vev [1] {\langle{#1}\rangle}

\begin{document}

\thispagestyle{empty}
\vspace{-1cm}
\rightline{DCPT-12/31}
\begin{center}
\vspace{1cm}
{\Large\bf  
Perturbative correlation functions \\
\vspace{.2cm}
of null Wilson loops and local operators }
\vspace{1.2cm}

{Luis F. Alday$^{a}$\footnote{alday@maths.ox.ac.uk}, Paul Heslop$^{b}$\footnote{paul.heslop@durham.ac.uk} and Jakub Sikorowski$^{c}$\footnote{j.sikorowski12@physics.ox.ac.uk}}\\
\vskip 0.6cm
{\em 
$^{a}$ Mathematical Institute, University of Oxford, 
   Oxford OX1 3LB, U.K.  \\
\vskip 0.08cm
\vskip 0.08cm $^{b}$ Department of Mathematical Sciences, Durham University, Durham DH1 3LE, U.K.\\
\vskip 0.08cm
\vskip 0.08cm
$^{c}$ Rudolf Peierls Centre for Theoretical Physics, University of Oxford, Oxford OX1 3NP, U.K.}
\vspace{.2cm}
\end{center}

\vspace{.4cm}
\begin{abstract} 
\noindent We consider the correlation function of a null Wilson loop with four edges and a local operator in planar MSYM. By applying the insertion procedure, developed for correlation functions of local operators, we give an integral representation for the result at one and two loops. We compute explicitly the one loop result and show that the two loop result is finite. 
\end{abstract}

\newpage
\setcounter{equation}{0} 
\setcounter{footnote}{0}
\setcounter{section}{0}

\setcounter{equation}{0} \setcounter{footnote}{0}
\setcounter{section}{0}
\vfuzz2pt 
\hfuzz2pt 

\setcounter{footnote}{0}


\setcounter{equation}{0}

\section{Introduction}

Correlation functions of gauge invariant local operators are the
natural observables of any conformal field theory. Over the last few
years, there has been rapid progress in the understanding/computation
of correlations functions of ${\cal N}=4$ SYM, see for instance
\cite{Eden:2011we,Eden:2012tu,Eden:2012fe}, and now explicit results, that would be impossible to obtain by standard Feynman diagram techniques, are available. 

Given an $n-$point correlation function $\langle {\cal O}(x_1)...{\cal
  O}(x_n) \rangle$ an interesting limit to consider is the one where
consecutive (after choosing a specific ordering)  distances became
null $x_{i,i+1}^2 \rightarrow 0$, at equal rate. It was argued in
\cite{Alday:2010zy} that in such a limit one obtains

\begin{equation}
\lim_{x_{i,i+1}^2 \rightarrow 0} \frac{\langle {\cal O}(x_1)...{\cal O}(x_n) \rangle}{\langle {\cal O}(x_1)...{\cal O}(x_n) \rangle_{\text{tree}}} = \langle W_{\text{adj}}^n[{\cal C}] \rangle
\end{equation}
where $W_{\text{adj}}^n[{\cal C}] $ is a Wilson loop in the adjoint representation, over the null polygonal path ${\cal C}$, with cusps at $x_i$. This relation is quite general and does not require the theory to be planar. If we focus on a planar theory, as we will do in this paper, then $\langle W_{\text{adj}}^n[{\cal C}] \rangle =\langle W_{\text{fund}}^n[{\cal C}] \rangle^2 $, the square of a Wilson loop in the standard fundamental representation.

One can also consider a generalization of the above limit, in which all distances but one became null. It was argued in \cite{Alday:2011ga}, see also \cite{Engelund:2011fg,Adamo:2011cd},  that in this limit one obtains

\begin{equation}
\label{partiallimit}
\lim_{x_{i,i+1}^2 \rightarrow 0} \frac{\langle {\cal O}(x_1)...{\cal
    O}(x_n) {\cal O}(y) \rangle}{\langle {\cal O}(x_1)...{\cal O}(x_n)
  \rangle} = \frac{ \langle W^n_{\text{adj}}[{\cal C}] {\cal O}(y)
  \rangle}{ \langle W^n_{\text{adj}}[{\cal C}]]\rangle}\ .
\end{equation}
On the right hand side we obtain the correlation function of a null Wilson loop with a local operator. This is a very interesting class of objects, in particular, they interpolate between a Wilson loop and a correlation function, and they are finite, since UV divergences in the numerator and denominator cancel out. The planar limit of a Wilson loop with operator insertions was discussed in detail in \cite{Engelund:2011fg}, where it was shown that $ \langle W^n_{\text{adj}}[{\cal C}] {\cal O}(y) \rangle \rightarrow 2 \langle W_{\text{fund}}^n[{\cal C}] \rangle  \langle W^n_{\text{fund}}[{\cal C}] {\cal O}(y) \rangle $. Hence, in the planar limit

\begin{equation}
\label{planar}
\lim_{x_{i,i+1}^2 \rightarrow 0} \frac{\langle {\cal O}(x_1)...{\cal O}(x_n) {\cal O}(y) \rangle}{\langle {\cal O}(x_1)...{\cal O}(x_n) \rangle} =2 \frac{ \langle W^n_{\text{fund}}[{\cal C}] {\cal O}(y) \rangle}{ \langle W^n_{\text{fund}}[{\cal C}]]\rangle}
\end{equation}

 In this paper we will focus on the simplest case, where the polygonal null Wilson loop has four edges, {\it i.e.} $n=4$.  In this case conformal symmetry implies:

\begin{equation}
\frac{\langle W^4(x_1,x_2,x_3,x_4) {\cal O}(y) \rangle}{\langle
  W^4(x_1,x_2,x_3,x_4)\rangle}=  \frac{|x_{13}
  x_{24}|^2}{\prod_{i=1}^4 |y-x_i|^2} F(\zeta)\ ,
\end{equation}
where $\zeta$ is the cross-ratio that can be constructed out of the location of the local operator $y$ and the location of the cusps $x_i$: 

\begin{equation}
\zeta = \frac{|y-x_2|^2 |y-x_4|^2 x_{13}^2}{|y-x_1|^2 |y-x_3|^2
  x_{24}^2}\ .
\end{equation}
Hence $F(\zeta)$ is a function of a single variable $\zeta$, in addition to the coupling constant $a=\frac{g^2N}{4\pi^2}$. From the definition of the cross-ratio and cyclic symmetry of the location of the cusps, we expect $F(\zeta)$ to have ``crossing'' symmetry:

\begin{equation}
\label{crossing}
F(\zeta)=F(1/\zeta)
\end{equation}
For the case of ${\cal O} ={\cal O}_{dil}$, the operator that couples to the dilaton ({\cal i.e.} the ${\cal N}=4$ action), this function was computed in \cite{Alday:2011ga} at leading order both in the weak and strong coupling expansions\footnote{In \cite{Alday:2011ga} the strong coupling result was found to be $F(\zeta)=c \frac{-\zeta}{3(1-\zeta)^3}(2(1-\zeta)+(\zeta+1)\log \zeta) \sqrt{\lambda}$, with $\lambda=4 \pi^2 a$. In the appendix we show that $c=-3/(4\pi^3)$ in order for (\ref{integratedF}) to be satisfied. As for the weak coupling result, one is to set $\hat{c}_{dil}=1/2$ in \cite{Alday:2011ga} , and further multiply by $1/4$, since   \cite{Alday:2011ga} used a non-standard convention for traces in the fundamental representation.} 

\begin{eqnarray}
F(\zeta) &=& -\frac{a}{4\pi^2}+...,~~~~~a \ll 1\\
F(\zeta)&=& \frac{\zeta}{(1-\zeta)^3}(2(1-\zeta)+(\zeta+1)\log \zeta) \frac{\sqrt{a}}{2\pi^2}+...,a \gg 1
\end{eqnarray}
we can see that both expressions satisfy the crossing symmetry (\ref{crossing}). The aim of the present paper is to compute $F(\zeta)$ to higher orders in perturbation theory. 

A related quantity, namely the four-point correlation function of the
stress-tensor multiplet,  has been extensively studied in the past as
well as
more recently and has now been explicitly computed at the integrand
level to 6 loops~\cite{oneTwo,ESS,Eden:2011we,Eden:2012tu}. This
multiplet, in particular, contains the chiral Lagrangian of ${\cal
  N}=4$ SYM. Computations of the correlator have made extensive  use
of  the method of Langrangian insertions. This method relies on the
observation that derivatives with respect to the coupling constant of
any correlation function can be expressed in terms of a  correlation function involving an additional insertion of the ${\cal N}=4$ SYM action. For instance,

\begin{equation}
a \frac{\partial}{\partial a} \langle {\cal O}(x_1)... {\cal O}(x_4)
\rangle = \int d^4 x_5  \langle {\cal O}(x_1)... {\cal O}(x_4) {\cal
  L}_{{\cal N}=4}(x_5)\rangle \ .
\end{equation}
This method is very powerful: by successive differentiation with
respect to the coupling, it allows one to express the $\ell-$loop correction for the four-point correlation in terms of the integrated tree-level correlation function with $\ell$ additional insertions of the ${\cal N}=4$ SYM Lagrangian.

From the discussion above it is clear that a particular limit of those integrands will produce the integrands for $\langle {\cal O}(x_1)... {\cal O}(x_4) {\cal L}_{{\cal N}=4}(x_5)\rangle$ in the particular null limit we are interested in. This will give integrand expressions for loop corrections to  $\langle W^4 {\cal L}_{{\cal N}=4}(x_5)\rangle$. 

In the next section we start by writing down those integral expressions. Then we compute the one-loop correction to $F(\zeta)$ (proportional to $a^2$) and show that the two-loop correction (proportional to $a^3$) is finite. This is to be expected, but it is far from obvious from the integral expressions, since each integral diverges as $\frac{1}{\epsilon^4}$ in dimensional-regularization.

Before proceeding, let us finish with a brief comment. The insertion procedure in particular implies an integral constraint on $F(\zeta)$, namely

\begin{equation}
\label{integratedF}
x_{13}^2 x_{24}^2 \int d^4 y \frac{ F(\zeta) }{\prod_{i=1}^4 |y-x_i|} = a \partial_a \log \langle W^4\rangle
\end{equation}
One can check that this equation is indeed satisfied by the leading
results at weak and at strong coupling, and we do so in the
appendix.

\section{Explicit results}

\subsection{General expressions and one-loop result}

Following \cite{Eden:2011we,Eden:2011yp,Eden:2011ku} we introduce

\begin{eqnarray*}
 \langle {\cal O}(x_1)...{\cal O}(x_4) \rangle&=&G_4 = \sum_{\ell=0}^\infty a^\ell G_4^{(\ell)}(1,2,3,4)\\
\langle {\cal O}(x_1)...{\cal O}(x_4) {\cal L}(x_5)\rangle&=& 1/4 \int d^4 \rho_5 G_{5;1} = 1/4\,\sum_{\ell=0}^\infty a^{\ell +1}  \int d^4 \rho_5 G_{5;1} ^{(\ell)}(1,2,3,4,5)
\end{eqnarray*}
here $\rho$ is a Grassmann variable, ${\cal O}$ is the lowest
component of the stress-tensor multiplet and ${\cal L}$ is the
component proportional to $\rho^4$. We define the 'tHooft coupling
constant $a=g^2N/(4\pi^2)$. The object we want to compute is then simply given by
\begin{equation}
\label{ratio}
 \frac{\langle {\cal O}(x_1)...{\cal O}(x_4) {\cal L}(x_5)\rangle}{ \langle {\cal O}(x_1)...{\cal O}(x_4) \rangle} =\frac{ \int d^4 \rho_5 G_{5;1} }{4G_4}
\end{equation}
Expressions for $G_4^{(\ell)}$ and  $G_{5;1} ^{(\ell)}$ (in terms of certain functions to be defined bellow), can be found in \cite{Eden:2011we}. In general, those depend on the insertion points, together with certain auxiliary harmonic variables $y_i$. In the null limit considered in this paper, however, the dependence on the  harmonic variables factors out, and goes away when taking the ratio (\ref{ratio}). In the null limit we obtain

\begin{align}
  G_{4}^{(\ell)}(1,2,3,4) \, &=\, \frac{1}{\ell!} \frac{2\,x_{13}^2x_{24}^2 }{(-4 \, \pi^2)^{\ell}} \,  G_4^{(0)}
  \, \int d^4x_{5} \dots d^4 x_{4+\ell} f^{(\ell)}(x_1, \dots,
  x_{4+\ell})\\
\int d^4 \rho_5  G_{5;1}^{(\ell)} \, &=  {\frac 8 {\ell!}}\frac{\,x_{13}^2x_{24}^2 }{(-4 \, \pi^2)^{\ell+1}} \,
  G_4^{(0)} \,  \int d^4x_{6} \dots d^4 x_{5+\ell}
  f^{(\ell+1)}(x_1, \dots, x_{5+\ell})\label{eq:3}
\end{align}
which is consistent with the insertion formula
\begin{align}
  a {\partial \over \partial a} G_{4}\, &= \, 1/4 \int d^4x_5
  \int d^4 \rho_5 G_{5;1}\ .
\end{align}
Finally we also need expressions for the $f$ functions. These have a
remarkably simple form~\cite{Eden:2011we}.
At $1,2,3$
loops these are given by\footnote{Note that the functions $f^{(\ell)}$
  are multiplied by the
  overall factor $(x_{12}^2x_{13}^2 x_{14}^2 x_{23}^2x_{24}^2
  x_{34}^2 )$ compared to the definition in~\cite{Eden:2011we}.}
\begin{align}\notag
 f^{(1)}(x_1,\ldots,x_5) &= {1 \over x_{15}^2 x_{25}^2x_{35}^2
   x_{45}^2}  \,,
\\ \label{f12}
 f^{(2)}(x_1,\ldots,x_6) &=  {\frac1{48}\sum_{\sigma \in S_6}x_{\sigma(1)
      \sigma(2)}^2 x_{\sigma(3) \sigma(4)}^2x_{\sigma(5)
      \sigma(6)}^2\over (x_{15}^2 x_{25}^2x_{35}^2
   x_{45}^2)( x_{16}^2 x_{26}^2x_{36}^2
   x_{46}^2) x_{56}^2}  \\
  f^{(3)}(x_1,\dots,x_7)&={{1 \over 20} \sum_{\sigma \in S_7} x_{\sigma_1\sigma_2}^4 x_{\sigma_3\sigma_4}^2 x_{\sigma_4\sigma_5}^2 x_{\sigma_5\sigma_6}^2 x_{\sigma_6\sigma_7}^2
   x_{\sigma_7\sigma_3}^2 \over (x_{15}^2 x_{25}^2x_{35}^2
   x_{45}^2)(x_{16}^2 x_{26}^2x_{36}^2
   x_{46}^2)(x_{17}^2 x_{27}^2x_{37}^2
   x_{47}^2)(x_{56}^2x_{57}^2 x_{67}^2)} \,.\notag
\end{align}
These functions satisfy certain symmetries. Upon multiplication by the
product of all external kinematic invariants $(x_{12}^2x_{13}^2 x_{14}^2 x_{23}^2x_{24}^2
  x_{34}^2 )$ and for generic (non-null-separated) points, these
  functions are completely symmetric under interchange of any two
  points and can be written as
  $\frac{P^{(\ell)}(x_1,...,x_{4+\ell})}{\prod_{1 \leq i < j \leq
      4+\ell}x_{ij}^2}$, where $P^{(\ell)}$ is a homogeneous
  polynomial in $x_{ij}^2$ of uniform weight $-(\ell-1)$ at each
  point. These properties hold at all loops in perturbation
  theory~\cite{Eden:2011we}. When taking the null limit the functions
  $f^{(\ell)}$ will have fewer terms, but some symmetries will be lost. 
  
Let us now consider the ratio (\ref{ratio}) order by order in perturbation theory

\begin{align}
   {\int d^4 \rho_5 G_{5;1} \over G_{4}} \notag
  &= \int d^4 \rho_5 \Bigg\{ a \Bigg[{G_{5;1}^{(0)}\over G_4^{(0)}}\Bigg] \ +\
  a^2 \,\Bigg[{G_{5;1}^{(1)}\over G_4^{(0)}}-{G_{5;1}^{(0)}\over
    G_4^{(0)}}{G_{4}^{(1)}\over G_4^{(0)}}\Bigg]\notag \\
  &+a^3 \,\Bigg[{G_{5;1}^{(2)}\over G_4^{(0)}}-{G_{5;1}^{(1)}\over
    G_4^{(0)}}{G_{4}^{(1)}\over G_4^{(0)}}-{G_{5;1}^{(0)}\over
    G_4^{(0)}}{G_{4}^{(2)}\over G_4^{(0)}}+{G_{5;1}^{(0)}\over
    G_4^{(0)}}\left({G_{4}^{(1)}\over G_4^{(0)}}\right)^2\Bigg]+\dots \Bigg\}
\end{align}
Hence, at leading order in perturbation theory (proportional to $a$) we find

\begin{align}
   \left({ \vev{W^4 {\cal L} } \over \vev{W^4}}\right)^{(0)}&= {a
      \over 8} {\int d^4 \rho_5 G_{5;1}^{(0)}\over G_{4}^{(0)}}= \frac{ a\,x_{13}^2x_{24}^2}{(-4 \, \pi^2)} \times  f^{(1)}(x_1, \dots, x_{5})\notag\\
   &= \frac{a }{(-4 \, \pi^2)}  {x_{13}^2x_{24}^2\over x_{15}^2
     x_{25}^2x_{35}^2x_{45}^2}\ ,
   \end{align}
which precisely agrees with the leading order result found in \cite{Alday:2011ga}. At next order we find
\begin{align}
    \left({ \vev{W^4 {\cal L} } \over \vev{W^4}}\right)^{(1)}& \ =\
    \frac{a^2 }{(-4 \, \pi^2)^{2}} \times x_{13}^2x_{24}^2 \times
   \Bigg[  \int d^4 x_6 f^{(2)}(x_1, \dots, x_{5},x_6) \notag\\
   &\qquad \qquad - 2x_{13}^2x_{24}^2 f^{(1)}(x_1, \dots, x_{5}) \int d^4 x_6 f^{(1)}(x_1, \dots, x_{4},x_6)\Bigg]\label{eq:4}
 \end{align}

In the
light-like limit the numerator of $f^{(2)}$ becomes simply
\begin{align}
&  \frac1{48}\sum_{\sigma \in S_6}x_{\sigma(1)
      \sigma(2)}^2 x_{\sigma(3) \sigma(4)}^2x_{\sigma(5)
      \sigma(6)}^2\notag\\&
    ={ x_{13}^2x_{24}^2x_{56}^2+ x_{15}^2
    x_{36}^2 x^2_{24} + x_{25}^2
    x_{46}^2 x^2_{13}+x_{35}^2
    x_{16}^2 x^2_{24}+x_{45}^2
    x_{26}^2 x^2_{13}}
\end{align}

When integrating over $x_6$ in (\ref{eq:4}) we recognize two kinds of contributions
\begin{eqnarray}
  F(1,2,3,4) = -{1\over 4\pi^2} \int d^4x_6 \frac{x_{13}^2x_{24}^2}{x_{16}^2 x_{26}^2 x_{36}^2 x_{46}^2} \\
  F(1,2,3,5) =  -{1\over 4\pi^2} \int d^4x_6 \frac{x_{13}^2x_{25}^2}{x_{16}^2 x_{26}^2 x_{36}^2 x_{56}^2} 
\end{eqnarray}
The first is the conformal massless box function, while the second is the two
mass hard box function (since $x_{51}$ and $x_{53}$ are not
null).\footnote{These integrals are of course infrared divergent and
  need regularisation. The combination of these integrals we consider
  below will be finite however and so we do not specify a regulator. In practise we will use dimensional regularisation (where the
  $x$'s are interpreted as dual momenta).} To be more precise,  we have
\begin{align}\label{eq:1}
& \left({ \vev{W^4 {\cal L} } \over \vev{W^4}}\right)^{(1)}=\frac{a^2 }{(-4 \, \pi^2)} \times  {x_{13}^2x_{24}^2\over x_{15}^2
   x_{25}^2x_{35}^2x_{45}^2} \\
& \times \Big(   F(1,2,3,5)+F(4,1,2,5)+F(3,4,1,5)+F(2,3,4,5)-F(1,2,3,4)\Big)\ .\nonumber
\end{align}
The first
($f^{(2)}$) term in~(\ref{eq:4}) contributes a similar expression with
all coefficients $+1$ 
whereas the second term in~(\ref{eq:4}) subtracts a term proportional to $2F(1,2,3,4)$ thus swapping the sign
of the last term.

The explicit expression for the box functions can be found for
instance in\cite{Bern:1994zx,Duplancic:2000sk}, where dimensional
regularization is used. Even though each box function is divergent,
the above combination is finite.  Furthermore this combination is
dual conformally invariant (see for example (2.23,2.22) of
\cite{Brandhuber:2009xz} for the divergences and
conformal variation of the box functions in dimensional regularization). Plugging the analytic expressions for the box functions and expanding
up to finite terms we obtain
\begin{align}
    \left({ \vev{W^4 {\cal L} } \over \vev{W^4}}\right)^{(1)} \ = \
    \frac{a^2 }{(-4 \, \pi^2)} \times  {x_{13}^2x_{24}^2\over x_{15}^2
     x_{25}^2x_{35}^2x_{45}^2} \times \left(-{1 \over 4}\right)\left( \log^2 \zeta+\pi^2 \right) 
\end{align}

\begin{equation}
  F(\zeta)= \frac{a^2}{(-4\pi^2)} \left(-{1 \over 4}\right) \left( \log^2 \zeta+\pi^2 \right)
\end{equation}
This result has homogeneous degree of transcendentality and the
correct symmetry $F(\zeta) = F(1/\zeta)$.

\subsection{Two-loop result}

At $O(a^3)$ we have
\begin{align}
  &    \left({ \vev{W^4 {\cal L} } \over \vev{W^4}}\right)^{(2)}\notag\\
  =& {1 \over 2}\frac{a^3 x_{13}^2x_{24}^2 }{(-4 \, \pi^2)^{3}} \times
  \int d^4 x_6 d^4 x_7 \times\notag\\
 & \times \Bigg[ f^{(3)}(x_1, \dots, x_{6},x_7) - 4x_{13}^2x_{24}^2 f^{(2)}(x_1, \dots, x_{6})f^{(1)}(x_1, \dots, x_4,x_{7}) \notag\\
   &\quad- 2x_{13}^2x_{24}^2 f^{(1)}(x_1, \dots, x_{5})f^{(2)}(x_1, \dots, x_4,x_6,x_{7})\notag\\
   &\quad+ 8 (x_{13}^2x_{24}^2)^2 f^{(1)}(x_1, \dots, x_{5})f^{(1)}(x_1, \dots, x_4,x_6)f^{(1)}(x_1, \dots, x_4,x_7) \Bigg]\label{eq:4a}
 \end{align}
The integrals which arise from this are a 2-mass pentabox,
2-mass (2 types) and massless double boxes, and products of massless and 2 mass
boxes. All of these are illustrated in the figures.
\begin{figure}[h!]
  \centering
  \setlength{\unitlength}{.45pt}
 \begin{picture}(220,233)
\put(0,0){\includegraphics[scale=.5]{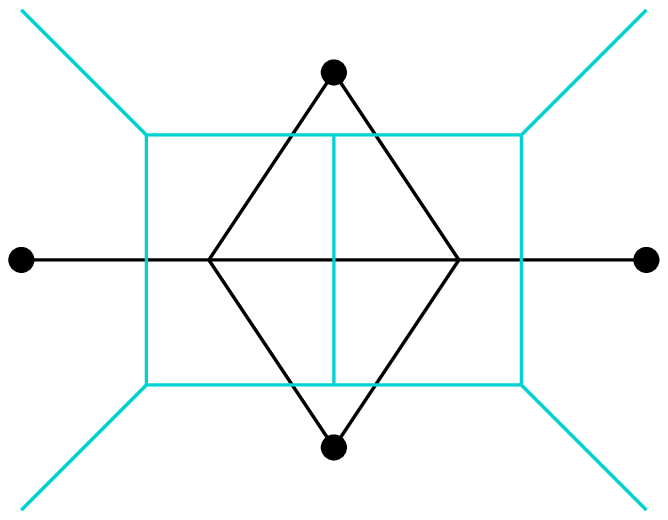}}
\put(108,222){\makebox(0,0)[lb]{\smash{$2$}}}
\put(1,150){\makebox(0,0)[lb]{\smash{$1$}}}
\put(106,78){\makebox(0,0)[lb]{\smash{$4$}}}
\put(219,150){\makebox(0,0)[lb]{\smash{$3$}}}
\put(97,10){\makebox(0,0)[lb]{\smash{$I_1$}}}
\put(83,51){\makebox(0,0)[lb]{\smash{${1\over 4}x_{13}^2 x_{24}^4$}}}
\end{picture}
\qquad
\begin{picture}(215,231)
\put(0,0){\includegraphics[scale=.45]{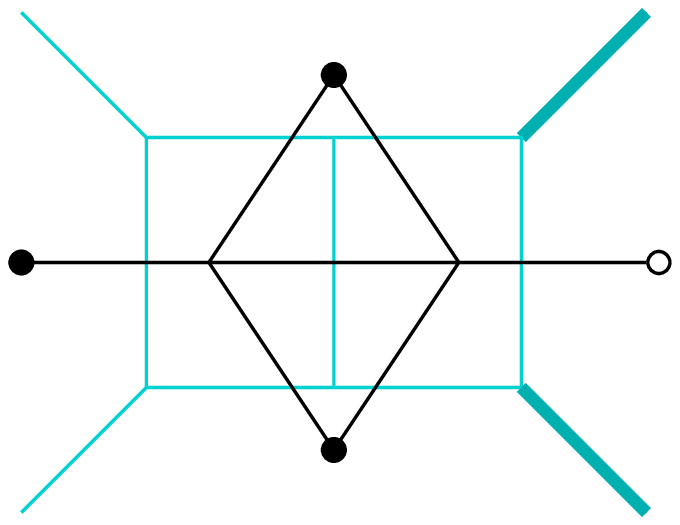}}
\put(214,149){\makebox(0,0)[lb]{\smash{ $5$}}}
\put(106,76){\makebox(0,0)[lb]{\smash{$1$}}}
\put(1,149){\makebox(0,0)[lb]{\smash{$2$}}}
\put(108,221){\makebox(0,0)[lb]{\smash{$3$}}}
\put(83,49){\makebox(0,0)[lb]{\smash{${1\over 2}x_{13}^4 x_{25}^2$}}}
\put(97,8){\makebox(0,0)[lb]{\smash{$I_2$}}}
\end{picture}
\qquad
\begin{picture}(213,228)
\put(0,0){\includegraphics[scale=.45]{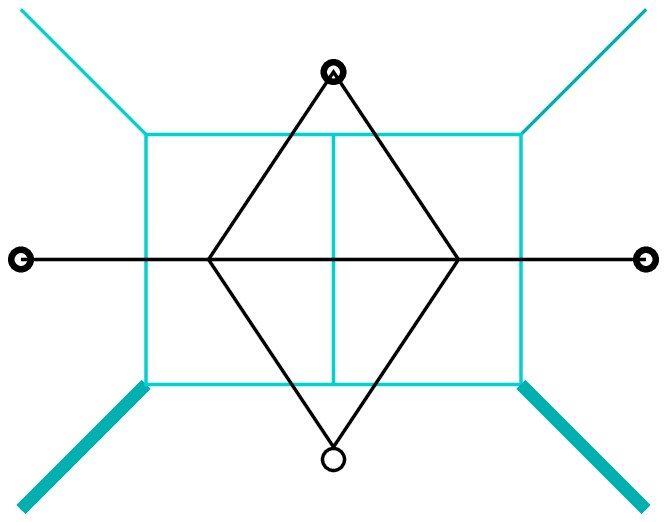}}
\put(103,77){\makebox(0,0)[lb]{\smash{$5$}}}
\put(212,148){\makebox(0,0)[lb]{\smash{$3$}}}
\put(107,218){\makebox(0,0)[lb]{\smash{$2$}}}
\put(1,149){\makebox(0,0)[lb]{\smash{$1$}}}
\put(80,49){\makebox(0,0)[lb]{\smash{${1\over 2}x_{13}^2 x_{25}^4$}}}
\put(94,8){\makebox(0,0)[lb]{\smash{$I_3$}}}
\end{picture}
  \caption{All contributing double box integrals at 2 loops with the corresponding numerator.
}
\end{figure}

\begin{figure}[h!]
   \setlength{\unitlength}{.45pt}
 \centering
     \begin{picture}(237,243)
       \put(0,0){\includegraphics[scale=.45]{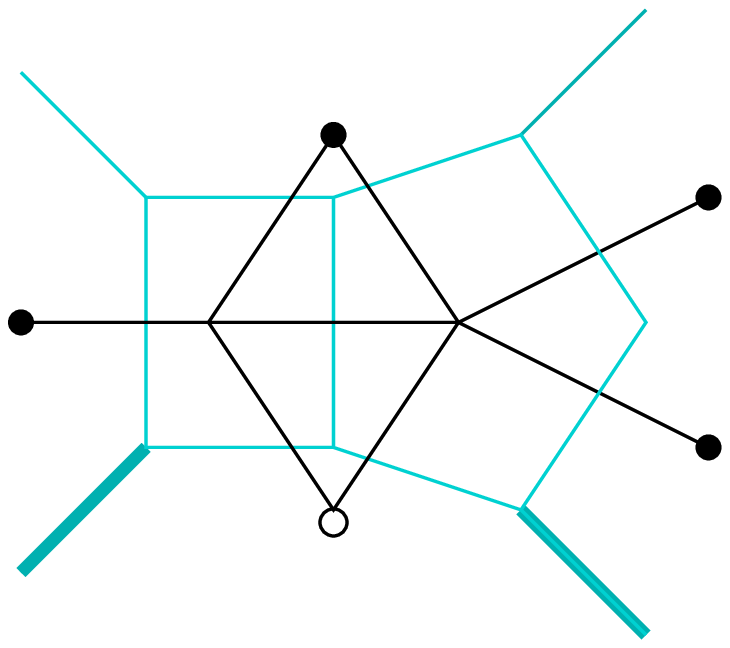}}
       \put(103,75){\makebox(0,0)[lb]{\smash{ $5$}}}
       \put(107,216){\makebox(0,0)[lb]{\smash{$2$}}}
       \put(1,147){\makebox(0,0)[lb]{\smash{$1$}}}
       \put(236,183){\makebox(0,0)[lb]{\smash{$3$}}}
       \put(231,110){\makebox(0,0)[lb]{\smash{$4$}}}
       \put(94,7){\makebox(0,0)[lb]{\smash{$I_4$}}}
       \put(142,161){\makebox(0,0)[lb]{\smash{$7$}}}
       \put(54,47){\makebox(0,0)[lb]{\smash{$x_{17}^2
             x_{24}^2x_{35}^2x_{25}^2$}}}
     \end{picture}
     \caption{The 2 mass pentabox which contributes at 2 loops with
       the corresponding numerator.}
   \end{figure}

    \begin{figure}[h!]
      \setlength{\unitlength}{.45pt}
     \begin{center}
      \begin{picture}(375,217)
        \put(0,0){\includegraphics[scale=.45]{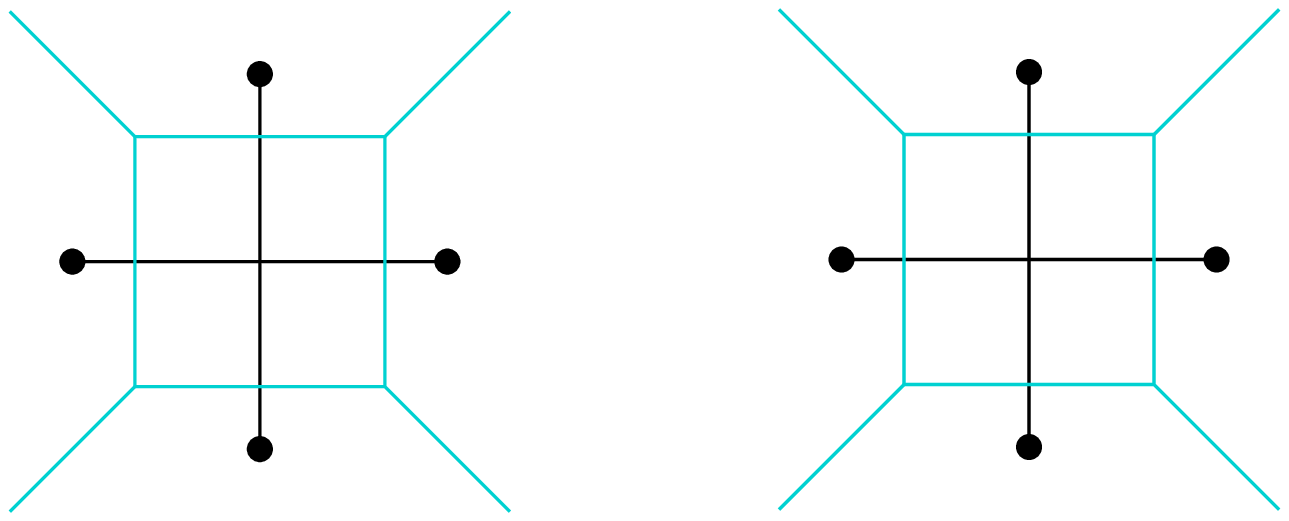}}
        \put(1,138){\makebox(0,0)[lb]{\smash{$1$}}}
        \put(144,137){\makebox(0,0)[lb]{\smash{$3$}}}
        \put(74,206){\makebox(0,0)[lb]{\smash{$2$}}}
        \put(70,65){\makebox(0,0)[lb]{\smash{$4$}}}
        \put(223,139){\makebox(0,0)[lb]{\smash{$1$}}}
        \put(366,138){\makebox(0,0)[lb]{\smash{$3$}}}
        \put(296,206){\makebox(0,0)[lb]{\smash{$2$}}}
        \put(293,66){\makebox(0,0)[lb]{\smash{$4$}}}
        \put(163,7){\makebox(0,0)[lb]{\smash{$I_5$}}}
        \put(175,130){\makebox(0,0)[lb]{\smash{$\times$}}}
        \put(135,39){\makebox(0,0)[lb]{\smash{${1\over 16}x_{13}^4
              x_{24}^4$}}}
      \end{picture}
    \end{center}

    \vskip.3cm

    \begin{center}
          \begin{picture}(375,217)
        \put(0,0){\includegraphics[scale=.45]{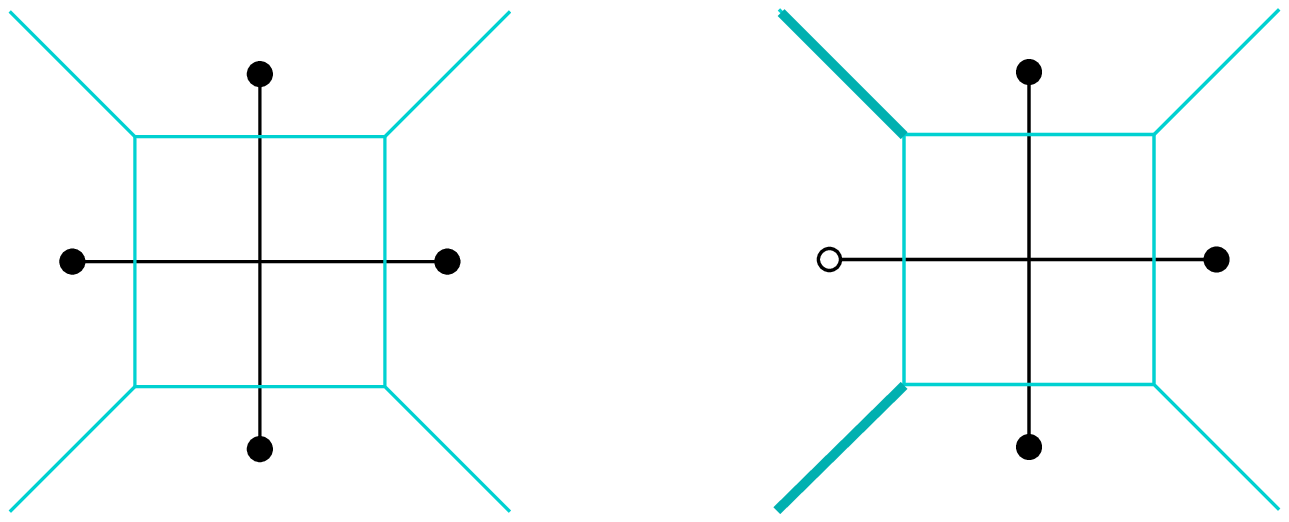}}
        \put(1,138){\makebox(0,0)[lb]{\smash{$1$}}}
        \put(144,137){\makebox(0,0)[lb]{\smash{$3$}}}
        \put(74,206){\makebox(0,0)[lb]{\smash{$2$}}}
        \put(70,65){\makebox(0,0)[lb]{\smash{$4$}}}
        \put(135,39){\makebox(0,0)[lb]{\smash{${1\over 2}x_{13}^2
              x_{24}^4x_{35}^2$}}}
        \put(163,7){\makebox(0,0)[lb]{\smash{$I_6$}}}
        \put(366,138){\makebox(0,0)[lb]{\smash{$3$}}}
        \put(296,206){\makebox(0,0)[lb]{\smash{$2$}}}
        \put(293,66){\makebox(0,0)[lb]{\smash{$4$}}}
        \put(175,130){\makebox(0,0)[lb]{\smash{$\times$}}}
        \put(223,139){\makebox(0,0)[lb]{\smash{$5$}}}
      \end{picture}
    \end{center}

    \vskip.3cm

    \begin{center}
            \begin{picture}(375,218)
        \put(0,0){\includegraphics[scale=.45]{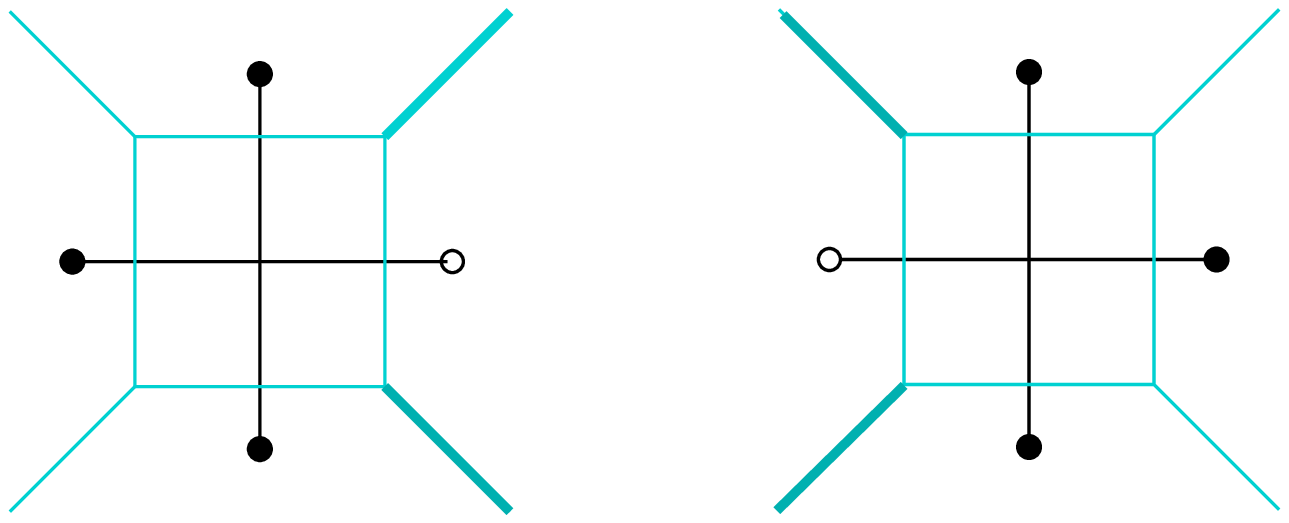}}
        \put(366,138){\makebox(0,0)[lb]{\smash{$3$}}}
        \put(296,206){\makebox(0,0)[lb]{\smash{$2$}}}
        \put(293,66){\makebox(0,0)[lb]{\smash{$4$}}}
        \put(175,130){\makebox(0,0)[lb]{\smash{$\times$}}}
        \put(223,139){\makebox(0,0)[lb]{\smash{$5$}}}
        \put(135,39){\makebox(0,0)[lb]{\smash{${1\over 4}x_{15}^2
              x_{24}^4x_{35}^2$}}}
        \put(1,138){\makebox(0,0)[lb]{\smash{$1$}}}
        \put(74,206){\makebox(0,0)[lb]{\smash{$2$}}}
        \put(70,65){\makebox(0,0)[lb]{\smash{$4$}}}
        \put(144,137){\makebox(0,0)[lb]{\smash{$5$}}}
        \put(163,7){\makebox(0,0)[lb]{\smash{$I_7$}}}
      \end{picture}
    \end{center}
      \caption{The 3 types of products of boxes which contribute at two loops  with the corresponding numerator.}
    \end{figure}

More specifically we have
\begin{align}
&  \int d^4x_6 d^4x_7{f^{(3)}(x_1, \dots, x_{6},x_7) \over f^{(1)}(x_1, \dots, x_{5})}=\sum_{\text{16 perms}}\Big(I_1 + I_2 + I_3 + I_4 + I_6+I_7
  \Big)\nonumber\\
&{x_{13}^2x_{24}^2 x_{13}^2x_{24}^2\over f^{(1)}(x_1, \dots, x_{5})}  \int d^4x_6 d^4x_7 f^{(2)}(x_1, \dots,
x_{6})f^{(1)}(x_1, \dots, x_4,x_{7})=\sum_{ \text{16 perms}}\Big(I_5
+{1\over 2}I_6\Big)\nonumber\\
&x_{13}^2x_{24}^2 \int d^4x_6 d^4x_7f^{(2)}(x_1, \dots,
x_4,x_6,x_{7})=\sum_{\text{16 perms}}\Big(I_1+I_5\Big)\nonumber\\
&(x_{13}^2x_{24}^2)^2  \int d^4x_6 d^4x_7 f^{(1)}(x_1, \dots,
x_4,x_6)f^{(1)}(x_1, \dots, x_4,x_7)=\sum_{\text{16 perms}} I_5
\end{align}
where the sum over 16 permutations indicates that we must sum over 16 permutations generated by cycling the external points $x_1,x_2,x_3,x_4$, parity ($x_1 \leftrightarrow x_4,\ x_2 \leftrightarrow x_3$) together with swapping the internal coordinates $x_6,x_7$. These permutations will not always produce a different integrand (for example $I_5$ is completely symmetric under all such permutations). We have divided by the corresponding symmetry factor in the definition of the integral (see figures).

Putting this all  together into~(\ref{eq:4a}) gives
\begin{align}
  &    \left({ \vev{W^4 {\cal L} } \over \vev{W^4}}\right)^{(2)}\notag\\
  &= {1 \over 2} {x_{13}^2 x_{24}^2 \over x_{15}^2x_{25}^2 x_{35}^2x_{45}^2} \times {a^3 \over (-4 \pi^2)^3} \times \sum_{\text{16 perms}} \Big(
  -I_1+I_2+I_3+I_4 +2I_5-I_6+I_7\Big)\ .\label{eq:5}
\end{align}

\subsection{Finiteness of the two-loop result}

We wish to check that the combination of (divergent) two-loop integrals~(\ref{eq:5}) is finite. To do this at the level of the integrand we have to first understand where the divergences come from, and then see if they cancel.
As is well-known there are two overlapping sources of infrared divergences for massless integrals, soft (when the internal momentum between two massless external legs vanishes) and collinear (when the integration momentum becomes collinear to a massless external leg)~\cite{Sterman:1995fz}. We make these divergences explicit in the current situation, by changing integration variables. Firstly we perform a transformation and Lorentz transformation to put $x_1=0$ and $x_2=(b/2,b/2,0,0)$. Next transform to the following variables $(\phi,\epsilon,\hat x_6)$ and $(\phi',\epsilon',\hat x_7)$ where 
\begin{align}
  x_6^\mu=\left(
    \begin{array}{c}
\phi (1+\epsilon^2)/2\\ \phi(1-\epsilon^2)/2\\ \phi \epsilon \hat x_6
\end{array}
\right)  \qquad   x_7^\mu=\left(
  \begin{array}{c}
\phi' (1+\epsilon'{}^2)/2\\ \phi'(1-\epsilon'{}^2)/2\\ \phi' \epsilon' \hat x_7
\end{array}
\right)\ .
\end{align}
Thus when $\epsilon\rightarrow 0$ we have $x_{61}$ collinear with $x_{12}$ whereas when $\phi\rightarrow 0$ we have $x_{71}\rightarrow 0$, so that the collinear and soft singularities occur when  $\epsilon=0$ and $\phi=0$ respectively. Similarly for $\epsilon'$ and $\phi'$ with $x_7$.
 Making this change of variables,
and focusing only on the potential divergences as $\epsilon,\epsilon',\phi,\phi'\rightarrow 0$, a generic two-loop integral takes the form
\begin{align}
 \text{finite} \times \int {d\epsilon \over \epsilon} {d\phi \over \phi} {d\epsilon' \over \epsilon'} {d\phi' \over \phi'} {\text{numerator}(\epsilon,\epsilon',\phi.\phi') \over x_{67}^2}\Big(1+O(\epsilon,\epsilon',\phi,\phi')\Big) \label{eq:7}
\end{align}
where
\begin{align}
  x_{67}^2=\phi^2\epsilon^2(1+\hat x_6^2)+\phi'{}^2\epsilon'{}^2(1+\hat x_7^2)+ \phi\phi'\Big(\epsilon^2+\epsilon'^2+2\epsilon\epsilon'\hat x_6\cdot\hat x_7\Big)\label{eq:6}
\end{align}
and where the numerator term is generically 4th order in
$\epsilon,\epsilon',\phi,\phi'$. So for example for the two loop
ladder diagram we have
\begin{align}
  \text{numerator}_{\text{ladder}}\sim \Big(\epsilon^2 \phi^2+ \epsilon'{}^2 \phi'{}^2\Big)\Big(1+O(\epsilon,\epsilon',\phi,\phi')\Big)\ ,
\end{align}
where in defining the integrand we always sum over permutations of the integration variables, giving the two terms here.
We then see that the integral diverges as
\begin{align}
  \text{ladder} \ &\sim\   \text{finite} \times \int {d\epsilon } {d\phi} {d\epsilon' \over \epsilon'} {d\phi' \over \phi'} {\epsilon \phi \over x_{67}^2}\Big(1+O(\epsilon,\epsilon',\phi,\phi')\Big) \\
  \ &\sim\ \text{finite} \times \int {d\epsilon } {d\phi} {d\epsilon' \over \epsilon'} {d\phi' \over \phi'} {1 \over \epsilon \phi }\Big(1+O(\epsilon,\epsilon',\phi,\phi')\Big)\ ,
\end{align}
where to get the second line we use that $x_{67}^2= \phi^2 \epsilon^2 (1+\hat x_6^2) + O(\epsilon',\phi')$ an approximation we can make, since there are poles in $\epsilon'$ and $\phi'$. We thus see the expected $\log^4$ divergence of the two-loop ladder.

Now, before addressing the case of interest, we consider another
interesting two-loop integral, namely the logarithm of the amplitude
at 2-loops. We know that this has a reduced infrared divergence, and it is interesting to see how this manifests itself at the level of the integrand. Again, performing the change of variables above we find that the numerator for the log of the amplitude takes the form
\begin{align}
  \text{numerator}_{\text{log of amplitude}}\sim \epsilon \phi
  \epsilon' \phi'\times\Big(1+O(\epsilon,\epsilon',\phi,\phi')\Big)
 \ .
\end{align}
Notice that this vanishes in any collinear ($\epsilon \rightarrow 0$)
or soft ($\phi \rightarrow 0$) limit in distinction with the ladder
diagram alone above. This shows that it must have a reduced divergence
compared to the ladder diagram. Indeed this simple fact (that the
numerator vanishes when the loop  integration becomes  collinear with
a massless external momentum) was used to great effect recently for
determining high-loop four-point amplitudes~\cite{Spradlin1,Spradlin2} and
correlation functions~\cite{Eden:2011we}. Here we go slightly further and consider the exact degree of divergence in this case.

Implementing the change of  integration variables, the log of the
amplitude takes the form
\begin{align}
  \text{log of amplitude} \ &\sim\   \text{finite} \times \int {d\epsilon } {d\phi} {d\epsilon'} {d\phi'} {1 \over x_{67}^2}\Big(1+O(\epsilon,\epsilon',\phi,\phi')\Big)\ .
\end{align}
To see the degree of divergence in this integral, it is useful to
change variable once more, and let $\epsilon' =\epsilon \alpha$,
$\phi'=\phi \beta$. Then the potential divergences occur at $\epsilon,
\phi, \alpha, \beta \rightarrow 0$ (also when $\alpha,\beta
\rightarrow \infty$ but we have symmetrized  the integration variables, allowing us to
concentrate on the former case). In these variables, from~(\ref{eq:6})
\begin{align}
  x_{67}^2 = \epsilon^2 \phi^2 \Big(1+\hat x_6^2+\alpha^2 \beta^2(1+\hat x_7^2)+ \beta \big(1+\alpha^2+2\beta \hat x_6\cdot\hat x_7\big) \Big)\\
  d\epsilon\,d\phi\,d\epsilon'\,d\phi' \ \rightarrow \
  d\epsilon\,d\phi\,d\alpha\,d\beta \times \epsilon \phi\ .
\end{align}
Then one can see there are no singularities when  $\alpha,\beta \rightarrow 0$ (unlike in the two-loop ladder case), and so the log of the amplitude takes the form
\begin{align}
  \text{log of amplitude} \ &\sim\   \text{finite} \times \int
  {d\epsilon } {d\phi}  {1 \over \epsilon \phi
  }\Big(1+O(\epsilon,\alpha,\phi,\beta)\Big)\ ,
\end{align}
and we identify the $\log^2$ divergence.

Finally then, we consider the case of interest, the two-loop integral defined in~(\ref{eq:5}). Making the change of variables we find that this time the numerator is of degree 6 in the $\epsilon, \epsilon', \phi. \phi'$ variables, in particular 
\begin{align}
  \text{numerator} \sim \epsilon \epsilon' \phi \phi' \big( A
  \,\epsilon \phi + B \,\epsilon' \phi'\big)\times \Big(1+O(\epsilon,\epsilon',\phi,\phi')\Big)\ .\label{eq:2}
\end{align}
for some finite $A,B$.

Let us then consider the degree of divergence of such an
integral. Plugging into~(\ref{eq:7}) and changing to the
$\alpha,\beta$ variables the numerator $\sim \epsilon^3 \phi^3 \alpha
\beta \big( a  + b \alpha \beta\big)$ and  there are then no poles at
all as $\epsilon, \phi, \alpha, \beta \rightarrow 0$ and thus the
integral is completely finite there. 

While perhaps not a completely rigorous proof of finiteness, the above argument
gives a strong indication that the above integral is
finite\footnote{Furthermore, numerical results are consistent with a cancelation of the two leading poles.}. Furthermore it provides an integrand-level criterion for
obtaining finite integrals: they must have numerators of the form
form~(\ref{eq:2}). Indeed implementing this criterion on  an
arbitrary linear combination of the integrals $I_1, \dots I_7$ gives
the unique solution~(\ref{eq:5}).

\section{Conclusions}

In this paper we considered the correlation function of a local
operator (the ${\cal N}=4$ Lagrangian) with a four cusped null Wilson
loop $\frac{\langle W^4{\cal O}(y) \rangle}{\langle W^4\rangle}$ in
perturbation theory. This correlation function is expected to be
finite, and conformal symmetry implies that the non-trivial dependence
is encoded in a function $F(\zeta)$ of a single cross-ratio and the
coupling constant. By using previous results on correlation functions, we computed $F(\zeta)$ at one-loop in perturbation theory, obtaining

\begin{equation}
F(\zeta) = -\frac{a}{4\pi^2} \left(1- \frac{1}{4}
  (\log^2\zeta+\pi^2)a+... \right)\ .
\end{equation}
Our result is consistent with crossing-symmetry $F(\zeta)=F(1/\zeta)$ and furthermore has the expected degree of transcendentality. Furthermore, we have given an integral representation for the two-loop contribution to $F(\zeta)$. This is given in terms of seven integrals, including double boxes and pentaboxes, plus permutations. Even though each contribution diverges as $1/\epsilon^4$ in dimensional regularization, we argue that this particular combination is finite. This claim is also supported by a numerical analysis of the integrals. We hope to come back in the future with a more detailed analysis, and hopefully an analytic answer, of the two-loop result. 

Finally, let us mention that the computation of $F(\zeta)$ should be simpler in certain limits. For instance, if the insertion point is null separated to one of the cusps (but not to the other) $\zeta$ vanishes (or becomes infinity), hence it should be possible to understand this limit in terms of the light-cone OPE for correlation functions. We hope to go back to this question in the future.

\bigskip

\section*{Acknowledgements}

PH gratefully acknowledges support from
STFC through the Consolidated Grant number ST/J000426/1 and of the ECT*, Trento.

\appendix

\section{Integration over the insertion point}

\subsection{Tree level}

In this appendix we will explicitly check that the normalization of the tree level result is consistent with \ref{integratedF}. At tree-level we have obtained $F=-\frac{a}{4\pi^2}$, hence, we should consider

\begin{equation}
I=-\frac{s~t~ a}{4 \pi^2} \int d^4 y \frac{1}{\prod_{i=1}^4 |y-x_i|} 
\end{equation}
where we have introduced $s=x_{13}^2$ and $t=x_{24}^2$. This is the usual massless scalar box function, and it has been computed in dimensional regularization, for instance, in \cite{Duplancic:2000sk}. At leading order in $\epsilon$ we obtain

$$I = -a \frac{1}{\epsilon^2}+...$$
We need to compare this with the divergent part of the right-hand-side of (\ref{integratedF}). This is

\begin{equation}
a \,\partial_a \log \langle W^4 \rangle_{div} = - \ell\, a^\ell\, \frac{\Gamma_{cusp}^{(\ell)}}{\epsilon^2}
\end{equation}
since $\Gamma_{cusp}=a+...$, we obtain the desired result.

\subsection{Strong coupling}

Let us now consider (\ref{integratedF}) at strong coupling. The unintegrated left hand side is the result of an integral over the world-sheet variables $(u,v)$ \cite{Alday:2011ga}

\begin{align}
  &x_{13}^2 x_{24}^2 \frac{F(\zeta)}{\prod_{i=1}^4|y-x_i|^2} \nonumber\\
  &= 2 c \int_{-\infty}^\infty \left(\frac{(\cosh u \sinh v)^{-1}}{1+y^2-2y_1 \tanh u-2 y_2 \tanh v+2 y_0 \tanh u \tanh v} \right)^4 du dv
\end{align}
where the world-sheet corresponds ends on the regular polygon with four edges. If we integrate over the world-sheet coordinates $(u,v)$ we obtain $F(\zeta)$ at strong coupling, quoted in the body of the text. On the other hand, we could also integrate over the location of the insertion point $y$. The integrals are quite elementary and we obtain
\begin{equation}
x_{13}^2 x_{24}^2 \int d^4 y \frac{F(\zeta)}{\prod_{i=1}^4|y-x_i|^2}= 2 c \frac{\pi^2}{6} \int_{-\infty}^\infty du dv
\end{equation}
If we set $c=-3/(4\pi^3)$, the right hand side coincides exactly with the action of the regular polygon with four edges found in \cite{Alday:2007hr}.

\end{document}